\documentclass{rsauthor}
\usepackage{amssymb}
\usepackage{graphicx}

\newcommand{\beq}{\vspace{-1.5mm}\begin{equation}}
\newcommand{\eeq}{\vspace{-1.5mm}\end{equation}}
\newcommand{\beqa}{\begin{eqnarray}}
\newcommand{\eeqan}{\end{eqnarray*}}
\newcommand{\beqan}{\begin{eqnarray*}}
\newcommand{\eeqa}{\end{eqnarray}}
\newcommand{\bra}[1]{\langle{#1}|}

\newcommand{\ket}[1]{|{#1}\rangle}

\newcommand{\id}{{\mathbf 1}}
\newcommand{\ip}[1]{\langle{#1}\rangle}

\newcommand{\eqr}[1]{equation (\ref{#1})}
\newcommand{\up}{\uparrow}
\newcommand{\dn}{\downarrow}
\newcommand{\tr}{{\rm Tr}}

\newcommand{\ri}{{\rm i}}

\begin{document}

\setlength{\textheight}{194mm}

\setcounter{page}{1065}
\jname{Proc. R. Soc. A (2012) {\bf 468},}%
\artnum{2011.0271}

\setlength{\oddsidemargin}{-1in}
\setlength{\evensidemargin}{-1in}

\setlength{\footskip}{0mm} \setlength{\voffset}{0mm}

\pagestyle{empty}\pagestyle{myheadings} \markright{J.A.Vaccaro\hfill
Particle-wave duality \hfill}

\title[Particle-wave duality]{Particle-wave duality: a dichotomy between symmetry and asymmetry}

\author[J.A. Vaccaro]{\sc By Joan A. Vaccaro}
\address{Centre for Quantum Computation and Communication Technology
(Australian Research Council),
         Centre for Quantum Dynamics, Griffith University, Brisbane, Queensland 4111, Australia}

\label{firstpage}

\maketitle \pagestyle{myheadings} \markright{J.A.Vaccaro\hfill Particle-wave
duality \hfill}

\begin{abstract}{particle-wave duality; complementarity;  information theory; group theory}
Symmetry plays a central role in many areas of modern physics.  Here we show
that it also underpins the dual particle and wave nature of quantum systems.
We begin by noting that a classical point particle breaks translational
symmetry whereas a wave with uniform amplitude does not.  This provides a
basis for associating particle nature with asymmetry and wave nature with
symmetry.  We derive expressions for the maximum amount of classical
information we can have about the symmetry and asymmetry of a quantum system
with respect to an arbitrary group. We find that the sum of the information
about the symmetry (wave nature) and the asymmetry (particle nature) is
bounded by $\log(D)$ where $D$ is the dimension of the Hilbert space. The
combination of multiple systems is shown to exhibit greater symmetry and thus
more wavelike character. In particular, a class of entangled systems is shown
to be capable of exhibiting wave-like symmetry as a whole while exhibiting
particle-like asymmetry internally. We also show that superdense coding can
be viewed as being essentially an interference phenomenon involving wave-like
symmetry with respect to the group of Pauli operators.
\end{abstract}

\section{Introduction}

The duality of particle and wave nature is one of the tenets of modern
quantum theory. Feynman summarised its importance by remarking that it
contains the {\it only} mystery of quantum theory (Feynman {\it et al.}
1963). Often the duality is rephrased in terms of Bohr's complementarity
principle (Bohr 1935) where particle nature is equated with well-defined
position and wave nature with well-defined momentum. In the last few decades
attempts have been made to quantify the duality more rigorously. For example
Wootters and Zurek (1979) formulated an inequality for a double slit
experiment that expresses a lower bound on the loss of path information for a
given sharpness of the interference pattern. The first experimental
realisation of distinct particle and wave properties of individual photons
was demonstrated by Grangier {\it et al.} (1986) using two different
experimental arrangements and a heralded single photon source. Scully {\em et
al.} (1991) explored the erasure of path information and the recovery of an
interference pattern using sub-ensembles conditioned on ancillary
measurements. A debate regarding the application of an uncertainty principle
ensured (see Wiseman 1998 and references therein). Later Englert (1996)
refined the mathematical representation of the duality by deriving an
inequality for a two-way interferometer that limits the distinguishability of
the outcomes of a path measurement and the visibility of the interference
pattern and made the distinction between a priori predictability and
distinguishability of the paths. Bj\"{o}rk and Karlsson (1998) then extended
the analysis to include quantum erasure. Barbieri {\it et al.} (2009)
recently verified Englert's duality relationship experimentally. The study of
the canonical position and momentum operators have been extended to general
canonically conjugate observables (Pegg {\it et al.} 1990) and their
properties explored using entropic uncertainty relations (Maassen \& Uffink
1988, Rojas Gonzalez {\it et al.} 1995) and other measures (Luis 2003). The
related study of the approximate simultaneous measurement of non-commuting
observables has also a long history (Arthurs \& Kelly 1965, Luis 2004, Ozawa
2004 and references therein). A different track has been to explore Bohr's
complementarity principle in terms of the mutually unbiased bases (MUB)
introduced by Schwinger (1960). The study of MUB is important for areas such
as discrete Wigner functions (Gibbons {\it et al.} 2004), quantum error
correction (Gottesman 1996), quantum cryptography (Miyadera \& Imai 2006) as
well as entangled systems (Kalev {\it et al.} 2009, Berta {\it et al.} 2010).
Kurzynski {\it et al.} (2010) have recently examined the physical meaning of
the operators associated with MUB for a spin-1 system.

Despite this work, there remain unexplored questions surrounding the mystery
of the particle-wave duality. As pointed out by Englert (1996), the notions
of wave and particle are borrowed from classical physics. An open question is
whether each classical notion should be represented by a {\it single} quantum
observable. In other words, can the problem be cast in terms of something
more general such as a symmetry of the quantum system? Instead of looking for
a pair of relevant observables could there not be a {\it set of observables
or operators} associated with each notion? Moreover, the MUB approach to this
problem yields complementary observables that tend to reflect mathematical
properties of the underlying Hilbert space rather than objects of direct
physical meaning (Kurzynski {\it et al.} 2010). This leads to the question of
whether the particle-way duality can be studied in a way that is {\it general
and yet retains a consistent physical basis}.  These are the key questions
that we address in this paper.

We begin by showing in \S2 that classical particle-like and wave-like
properties have natural definitions in terms of a symmetry group. In \S3 we
differentiate between two sets of operations on the quantum system according
to their effect on the symmetry of the system. This allows us to associate a
set of operators that manipulate only particle-like properties (or asymmetry)
of the system and another set that manipulate only the wave-like properties
(or symmetry). We then define a convenient measure of the degree to which a
state of the system exhibits particle or wave properties based on the ability
to encode information in the system using the corresponding set of
operations. A duality relation between the symmetry and asymmetry is derived
in \S4. In \S5 we extend the analysis to composite systems. Applications of
the formalism are given in \S6 and we end with a discussion in \S7.  A
preliminary version of this work was presented in Vaccaro 2006.

\section{Symmetry of particles and waves \label{sec:symmetry}}

\subsection{Particle-like and wave-like properties \label{sec:part and wave props}}

\begin{figure}  
\begin{center}
\includegraphics[width=12cm,height=3.3cm]{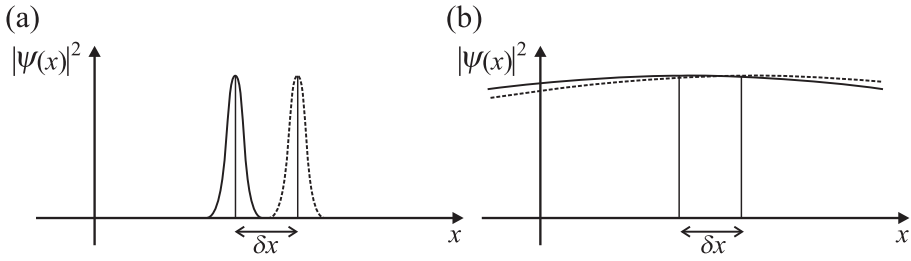}
\end{center}
\caption{Spatial translations for (a) a narrow particle-like wave function
and (b) a broad wave-like wave function. Solid curves represent the original
functions and dashed curves the displaced versions for a translation of
$\delta x$ to the right. The wave functions are not normalised. \label{fig1:
spatial translations}}
\end{figure}    

Classical particles and waves respond in distinct ways to a spatial
translation: the position of a classical point particle is displaced whereas
the amplitude function of a uniform classical wave is invariant. These
responses provide a basis for defining analogous wave and particle properties
of quantum systems. Consider first a quantum system whose wave function
comprises a relatively narrow peak in the position representation, as
illustrated in figure \ref{fig1: spatial translations}(a). A spatial
translation of $\delta x$ along the $x$ axis of sufficient magnitude can
completely displace the system so that its wave function $\psi(x)$ is mapped
to an orthogonal wave function $\psi(x-\delta x)$.  For example, the overlap
$\int \psi^*(x)\psi(x-\delta x) dx$ is negligible for Gaussian wave functions
of the kind $\psi(x)\propto \exp(-x^2/4\sigma^2)$ with $\sigma\ll \delta x$.
Thus quantum systems with relatively narrow wave functions in the position
representation behave as classical particles under spatial translations, as
one would expect. Next, consider a system whose wave function is delocalised
in the position representation so that the position probability density ${\rm
Pr}(x)=|\psi(x)|^2$ is relatively ``flat'', as illustrated in figure
\ref{fig1: spatial translations}(b). Such wave functions are relatively
invariant to spatial translations and so the system behaves as a classical
wave under spatial translations. For example ${\rm Pr}(x-\delta x)\approx{\rm
Pr}(x)$ for Gaussian wave functions of the kind
$\psi(x)\propto\exp(-x^2/4\sigma^2)$ with $\sigma\gg \delta x$. Particle-like
and wave-like properties of quantum systems can therefore be distinguished by
whether the system is displaced or invariant, respectively, to spatial
translations.

The invariance of a system to a given set of operations represents a
particular symmetry of the system.  In the case here wave-like properties
represent the {\it symmetry} of a quantum system with respect to the group of
spatial translations. Conversely, any displacement of a system under a
spatial translation is a lack of this symmetry. Thus particle-like properties
represent the {\it asymmetry} of a quantum system with respect to the group
of spatial translations.  In the following sections we generalise this
concept by associating wave and particle properties with symmetry and
asymmetry, respectively, for arbitrary groups.

\subsection{Symmetry and interference  \label{sec:sym and interference}}

However, we should not loose sight of the fact that for a system to be
regarded as wave-like it must have an ability to produce interference in an
interferometer of some kind. Symmetry, in contrast, is an intrinsic property
of the system and can be quantified without reference to the details of any
interferometer. We can, however, relate symmetry directly to interference in
the following way.  Taking the double slit experiment as the prototypical
interferometer, we note that the positions of interference fringes on the
screen can be changed by introducing a relative phase shift at the slits.
This property allows an interferometer to be used as a communication channel
between a sender at the slits and a receiver at the screen. The sender can
encode a message by modifying the relative phase between the slits and the
receiver can faithfully decode the message by observing the position of the
fringes on the screen. In the next section we quantify the degree of symmetry
of a system by its capacity to carry information in a likewise manner. In
other words, an interferometer can be viewed as a communication channel and
symmetry is measured in terms of information capacity. This information
theoretic link between symmetry and interference applies to arbitrary
systems. More will be said about this later.

Interference and symmetry share another common feature that is worth
mentioning. Interference is only ever seen in a statistical sense and its
full characterisation requires infinitely many observational events.  For
example, to see the wave-like character of an electron, we need a beam of
electrons rather than a single one. This implies the wave-like character that
we infer from interference has a similar statistical meaning. Likewise, the
measure of symmetry that we use below is in terms of its capacity to carry
information, and that capacity is derived by considering the statistics of
random messages in the limit of infinitely many messages. So both have a
statistical character and represent infinite ensembles. Nevertheless, we
shall refer to the wave-like character and the symmetry of a single system.

Also while interference is typically viewed as occurring in coordinate space,
there have been studies of interference in momentum space. For example, Rauch
(1993) argued that in neutron interference experiments, when the path
difference is sufficiently long that the neutron wave packets no longer
overlap in space and the spatial interference has vanished, there can be
persistent interference in the momentum representation in one of the output
paths of the interferometer. In a similar vein, Pitaevskii and Stringari
(1999) have shown that optically probing two spatially separated cold atomic
gases can reveal interference in the momentum representation. Remarkably
P\'{e}rez Prieto {\it et al.} (2001) have shown that in the collision of a
wave packet with a potential barrier there can be suppression of a particular
momentum value and enhancement of others due to the interference between the
transmitted and reflected parts of the wave packet. Recently Ruschhaupt {\it
et al.} (2009) took this a step further and devised a momentum-space
interferometer using a trapped cold Bosonic atomic gas. Their proposal is to
phase imprint part of the gas cloud using a detuned laser. The imprinting
ensures that the wave function in the momentum representation comprises a sum
of two terms which interfere. Significantly, changing the amount of phase
imprinting results in shifting the position of a node in the momentum
distribution. These studies show that interference fringes may be present in
the momentum representation and not the spatial representation, and in doing
so they call for a broader interpretation of what constitutes wave-like
character. But this does not pose a problem for us here. Indeed the arguments
given above for symmetry and interference in configuration space also hold in
momentum space provided that the symmetry in question is the invariance to
momentum translations as opposed to spatial ones. In other words, wave-like
character is ascribable to the symmetry of momentum space. This highlights
the generality of our analysis in that it can capture the wave-like character
of arbitrary symmetries.

\subsection{Review of symmetry and asymmetry \label{sec:review sym and asym}}

Before beginning our analysis in detail it will be useful to collect a number
of relevant definitions and results. These relate to the symmetry and
asymmetry of states with respect to a finite or compact Lie group.  Let the
group be $G=\{g_1, g_2, \ldots\}$ and have the unitary representation $\{\hat
T_g: g\in G\}$ on the system's Hilbert space $H$.  We can borrow pertinent
results about symmetry from the study of superselection rules (SSRs).  A
review of recent work on SSRs in the context of quantum information theory
has been given by Bartlett {\it et al.} (2007). A state of the system is
symmetric with respect to $G$ if it satisfies
\beq \label{eq: G rho=rho}
   {\cal G}[\hat\rho] = \hat\rho
\eeq
where the ``twirl'' of $\hat\rho$ is defined as (Bartlett \& Wiseman 2003)
\beq  \label{eq: twirl}
   {\cal G}[\hat\rho] \equiv \frac{1}{n_G}\sum_{g\in G}
        \hat T_g \hat\rho\hat T^\dagger_g\ ,
\eeq
$n_G$ is the order of $G$ and $\hat \rho$ is the system's density operator.
Throughout this paper we use the notation for a finite group when referring
to an arbitrary group. The equivalent results for a compact Lie group are
easily found by appropriate modification of notation, for example, by
replacing the averaged sum in \eqr{eq: twirl} with an integral with respect
to an invariant measure on the group. We shall call states satisfying
\eqr{eq: G rho=rho} symmetric states. A symmetric state is unchanged by the
actions of the group. Indeed \eqr{eq: G rho=rho} implies
\beq
   \hat T_g\hat \rho \hat T_g^\dagger = \hat T_g {\cal G}[\hat\rho] \hat
   T_g^\dagger={\cal G}[\hat\rho] =\hat\rho
\eeq
for $g\in G$.  In contrast, a state exhibits asymmetry with respect to $G$ if
\beq  \label{eq: defn asym}
    \hat T_g\hat \rho \hat T_g^\dagger\ne \hat\rho
\eeq
for any $g\in G$.  In this case we shall refer to $\hat\rho$ as an asymmetric
state or as having asymmetry with respect to $G$.  A maximally-asymmetric
pure state is a pure state for which ${\cal G}[\hat\rho]$ is maximally mixed.

The degree to which a state is symmetric is given by the entropic measure
(Vaccaro {\it et al.} 2008)
\beq    \label{eq: W_G}
  W_G(\hat\rho) = \log(D)-S({\cal G}[\hat\rho])\ .
\eeq
where $S(\hat\varrho)=-{\rm Tr}[\hat\varrho\log(\hat\varrho)]$ is the von
Neumann entropy of $\hat\varrho$ and $D$ is the dimension of the system's
Hilbert space.  In the following we adopt the convention of information
theory and use the binary logarithm for $\log(\cdot)$. The measure
$W_G(\hat\rho)$ has been called the {\it symmetry} of the state $\hat\rho$
with respect to $G$. It has a maximum value of $\log(D)$ for a symmetric pure
state and a minimum of zero for a maximally-asymmetric pure state. Similarly,
the {\it asymmetry} of $\hat\rho$ with respect to $G$ has been defined as the
lack of symmetry as (Vaccaro {\it et al.} 2008)
\beq  \label{eq: A_G}
   A_G(\hat\rho)=S({\cal G}[\hat\rho])-S(\hat\rho)\ .
\eeq
The asymmetry $A_G(\hat\rho)$ represents the ability of a system in state
$\hat\rho$ to act as a reference to alleviate the effects of the SSR.

It will be useful to have states with maximal asymmetry for exploring maximal
particle-like properties. A system that is in a state with some asymmetry can
be used as a reference to partially break the symmetry represented by the
group $G$.  For example, an object which lacks spherical symmetry can be used
as a reference for orientation; the less spherical symmetry it has the better
reference it is. Spherical symmetry is broken for systems that include the
object. A quantum system can completely break the symmetry represented by the
group $G$ if its Hilbert space contains ``reference states'' for $G$ of the
form (Kitaev {\it et al.} 2004)
\beq   \label{eq: reference state |g>}
    \ket{g}=\sum_{q,i,a}\sqrt{\frac{n_q}{n_G}}D^{(q)}_{i,a}(g)\ket{q,i,a}
\eeq
for $g\in G$.  Here $q$ uniquely labels an irreducible representation
\beq
    D^{(q)}_{k,j}(g)=\ip{q,k,a|\hat T_g|q,j,a}
\eeq
of $G$ whose dimension is $n_q$, $a$ indexes different copies of each
irreducible representation, $\ket{q,i,a}$ are the basis states for which
$\hat T_g$ is block diagonal, the number of copies of each representation
equals the dimension of the representation $n_q$, the sum over $q$ ranges
over all irreducible representations and the dimension $D$ of the Hilbert
space $H$ spanned by the basis states $\ket{q,i,a}$ is such that $D=n_G$. The
reference states $\ket{g}$ have the property that their ``orientation'' is
changed by the action of the group, viz. $\hat T(g')\ket{g}=\ket{g'\circ g}$
and $\ip{g|g'}=\delta_{g,g'}$, and so the set of reference states $\{\ket{g}:
g\in G\}$ forms an orthonormal basis for $H$. They completely break the
symmetry of $G$ in the sense that the action of the group $\hat T(g')\ket{g}$
for $g'\in G$ generates a set of mutually orthogonal states.  In other words
the reference states $\ket{g}$ give a distinct orientation for each element
of the group. We shall refer to systems of this kind as capable of completely
breaking the symmetry of $G$.

We will also need to distinguish between different kinds of operators based
on their effect on the symmetry or asymmetry.  We already have one set
representing the group $G$ and can use it to define another. A unitary
operator $\hat U$ is called $G$-invariant if
\beq \label{eq: G invariant op}
   \hat U \hat T_g \hat \rho \hat T_g^\dagger\hat U^\dagger
      =\hat T_g \hat U \hat \rho \hat U^\dagger\hat T_g^\dagger
\eeq
for all $g\in G$ and all states $\hat\rho$.  It has been shown that
$G$-invariant operations cannot increase the asymmetry of a state either
individually or on average (Vaccaro {\it et al.} 2008).

We want to know how much information about the symmetry and asymmetry of a
quantum system with respect to $G$ is associated with the knowledge of its
density operator $\hat\rho$. For this we need to see how much information can
be encoded using the symmetry and asymmetry of the system.

\section{Information capacity of symmetry and asymmetry \label{sec: info
capacity}}

\subsection{Asymmetry}

We first examine how any asymmetry of $\hat\rho$ can be used to send
information. A state which is asymmetric is transformed by the actions of the
group $G=\{g\}$ to a different state. This means that the transformation
\beq
    \hat\rho\mapsto\hat\rho_g = \hat T_g\hat\rho \hat T_g^\dagger
\eeq
for $g\in G$ can carry information.  Moreover, it is straightforward to show
that ${\cal G}[\hat T_g\hat\rho\hat T_g^\dagger]={\cal G}[\hat\rho]$ and so
\beq  \label{eq: W_G (rho_g) = W_G (G[rho]) = W_G(rho)}
    W_G(\hat T_g\hat\rho\hat T_g^\dagger)
       =W_G(\sum_{g\in G}p_g\hat T_g\hat\rho\hat T_g^\dagger)=W_G(\hat\rho)
\eeq
for all $g\in G$, where $\{p_g\}$ is a probability distribution.  This means
that the operators $\hat T_g$ do not change the value of the symmetry
$W_G(\hat\rho)$ either individually or on average. Their use will ensure that
the encoding involves only the asymmetry of the system.

We imagine an information theoretic scenario where one party, Alice, sends
another, Bob, information encoded in the transformed states $\hat\rho_g$.
Specifically Alice prepares the system in the transformed state $\hat\rho_g$
for $g\in G$ with probability $p_g$ and sends it to Bob. We would like Alice
to encode the maximum amount of information in the asymmetry of $\hat\rho$.
If the probabilities $p_g$ were not all equal then the averaged state
prepared by Alice $\hat\rho_{\rm av}=\sum_g p_g \hat T_g\hat\rho \hat
T_g^\dagger$ might well be asymmetric according to \eqr{eq: defn asym} in
that
\beq
   \hat T_g\hat\rho_{\rm av}\hat T_g^\dagger\ne\hat\rho_{\rm av}
\eeq
for some $g\in G$. In this case the maximum amount of information is not
guaranteed to be encoded because any asymmetry of $\hat\rho_{\rm av}$ could
be used to encode additional information using the $T_g$ operators. In
contrast, if the probabilities are all equal the averaged prepared state is
given by $\hat\rho_{\rm av}={\cal G}[\hat\rho]$, as defined in \eqr{eq:
twirl}, which is symmetric because it is invariant to the actions of the
group, i.e.
\beq
   \hat T_g\Big({\cal G}[\hat\rho]\Big)\hat T_g^\dagger = {\cal G}[\hat\rho]\ \ {\rm for\ }g\in
   G\ .
\eeq
No further information can be encoded in ${\cal G}[\hat\rho]$ using the $T_g$
operators and so the encoding is maximal. We therefore stipulate that Alice
prepares each state $\hat\rho_g$ with equal probability, $p_g=1/n_G$ so that
the averaged prepared state is ${\cal G}[\hat\rho]$.

On receiving the system, Bob makes a measurement to estimate the value of the
parameter $g$. In particular, let Bob make the measurement ${\cal M}$
described by the Kraus operators $\hat M_k$ satisfying $\sum_k\hat
M_k^\dagger\hat M_k=\hat\id$ (Kraus 1983). Bob obtains result $k$ with
probability $P_{k,g}=\tr(\hat M_k^\dagger\hat M_k\hat\rho_g)$ for the
prepared state $\hat\rho_g$. The mutual information shared by Alice and Bob
about their respective indices $g$ and $k$  is given by (Schumacher {\it et
al.} 1996)
\beq
  I({\cal M}:{\cal G})=H(\{p_g\})-\sum_k P(k) H(\{P(g|k)\})
\eeq
where $H(\{q_n\})=-\sum_n q_n\log(q_n)$ is the Shannon entropy of the
probability distribution $\{q_n\}$, $P(k)=\sum_g p_g P(k|g)$ is the average
probability of getting result $k$ and $P(g|k)=p_g P(k|g)/P(k)$ is the
conditional probability that Alice prepared the state $\hat\rho_g$ given that
Bob obtained measurement outcome $k$.

We define the information capacity, $I_{\rm asym}(\hat\rho)$, of the
asymmetry of $\hat\rho$ as the accessible information about the value of $g$
carried by the ensemble of states $\{\hat\rho_g:g=1,2,\ldots,n_G\}$ sent to
Bob. This is given by the maximum of $I({\cal M}:{\cal G})$ over all possible
measurements ${\cal M}$ by Bob,
\beq
  \label{I_asym defn}
   I_{\rm asym}(\hat\rho)=\max_{\cal M} I({\cal M}:{\cal G})\ ,
\eeq
and it is bounded above by Holevo's theorem (Schumacher {\it et al.} 1996,
Ruskai 2002):
\beq
  \label{N_P_bound}
   I_{\rm asym}(\hat\rho)\le S({\cal G}[\hat\rho])-\frac{1}{n_G}\sum_{g\in G}
   S(\hat\rho_g)\ .
\eeq
As $\hat T_g$ is unitary, $S(\hat\rho_g)=S(\hat T_g\hat\rho\hat
T_g^\dagger)=S(\hat\rho)$ for all $g\in G$, and we find \eqr{N_P_bound}
becomes (Vaccaro 2006, see also Gour {\it et al.} 2009)
\beq
  \label{eq: I_as bound}
   I_{\rm asym}(\hat\rho)\le S({\cal G}[\hat\rho])-S(\hat\rho) \ .
\eeq
Comparing the right side with \eqr{eq: A_G} shows that the information
capacity $I_{\rm asym}(\hat\rho)$ is bounded by the entropic measure of the
asymmetry $A_G(\hat\rho)$.

There is another way of interpreting this result.  The value of $g$ indexes
different ``orientations'' of the system due to its asymmetry. Information
about $g$ is therefore information about the asymmetry of system.  There is a
total of $I_{\rm asym}(\hat\rho)$ bits of classical information about the
asymmetry being sent to Bob and this is the maximum amount possible. We
conclude that {\it Alice's knowledge of the original state $\hat\rho$
represents $I_{\rm asym}(\hat\rho)$ bits of classical information about the
asymmetry.}

\subsection{Symmetry  \label{subsec: Info capacity - sym}}

We now consider the analogous information theoretic scenario that uses the
symmetry part of the state $\hat\rho$.  To avoid using the asymmetry, we want
the set of operations used for the encoding to leave the value of the
asymmetry $A_G(\hat\rho)$ unchanged both on individual application and on
average. This requirement is the complement of \eqr{eq: W_G (rho_g) = W_G
(G[rho]) = W_G(rho)}. A $G$-invariant unitary operator $\hat U$ defined in
\eqr{eq: G invariant op} has the properties that $S({\cal G}[\hat
U\hat\rho\hat U^\dagger])=S(\hat U{\cal G}[\hat\rho]U^\dagger)=S({\cal
G}[\hat\rho])$ and $S(\hat U\hat\rho\hat U^\dagger)=S(\hat\rho)$, and so it
does not change the asymmetry of $\hat\rho$, i.e $A_G(\hat U\hat\rho\hat
U^\dagger)=A_G(\hat\rho)$.  Let Alice use a subset of $G$-invariant unitary
operators $U=\{\hat U_1, \hat U_2, \ldots, \hat U_N\}$ to encode information
in the system by preparing the state
\beq  \label{eq: rho_j}
   \hat\rho_j=\hat U_j\hat\rho\hat U_j^\dagger
\eeq
with probability $q_j$.  The choice of the subset must not change the
asymmetry of the state on average and so
\beq  \label{eq: A_G (U[rho]) = A_G(rho)}
    A_G(\hat\rho_{\rm av})=A_G(\hat\rho)
\eeq
where $\hat\rho_{\rm av}=\sum_j q_j\hat\rho_j$.  Equation (\ref{eq: A_G
(U[rho]) = A_G(rho)}) ensures that the encoding involves only the symmetry of
the system. Moreover, we want Alice to encode the maximum amount of
information in the symmetry. This means that the averaged state after the
encoding should have no symmetry, i.e. $W_G(\hat\rho_{\rm av})=0$ which, from
\eqr{eq: W_G}, implies
\beq  \label{eq: max info in symmetry}
    S({\cal G}[\hat\rho_{\rm av}])=\log(D)\ .
\eeq

The system is then sent to Bob who makes a measurement to estimate the value
of the parameter $j$. Let Bob's measurement be described as before. The
mutual information shared by Alice and Bob about their respective indices $j$
and $k$  is given by
\beq
  I({\cal M}:{\cal U})=H(\{p_j\})-\sum_k P(k) H(\{P(j|k)\})
\eeq
where ${\cal U}$ represents the encoding, $P(k)=\sum_j p_j P(k|j)$ is the
average probability of getting result $k$ and $P(j|k)=p_j P(k|j)/P(k)$ is the
probability that Alice prepared the state $\hat\rho_j$ given that Bob
obtained measurement outcome $k$.  We define the information capacity,
$I_{\rm sym}(\hat\rho)$, of the symmetry of $\hat\rho$ as the accessible
information about the value of $j$ carried by the ensemble of states
$\{\hat\rho_j:j=1,2,\ldots,N\}$ sent to Bob. This is given by the maximum of
$I({\cal M}:{\cal U})$ over all possible measurements ${\cal M}$ by Bob,
\beq  \label{eq: I_sym defn}
   I_{\rm sym}(\hat\rho)=\max_{\cal M} I({\cal M}:{\cal U})\ .
\eeq
From Holevo's theorem we find
\beq
  I_{\rm sym}(\hat\rho)\le S(\hat\rho_{\rm av})
      -\sum_{j} q_{j}S(\hat\rho_{j})\ ,
\eeq
and as the encoding is unitary $S(\hat\rho_{j})=S(\hat\rho)$ and so
\beq   \label{eq: I_sym 1st version}
  I_{\rm sym}(\hat\rho)\le S(\hat\rho_{\rm av})
     -S(\hat\rho)\ .
\eeq
Equation (\ref{eq: A_G (U[rho]) = A_G(rho)}) can be written as
\beq
   S({\cal G}[\hat\rho_{\rm av}])-S(\hat\rho_{\rm av})=S({\cal
   G}[\hat\rho])-S(\hat\rho)\ ,
\eeq
which can be rearranged to
\beq
        S({\cal G}[\hat\rho_{\rm av}])-S({\cal G}[\hat\rho])
          = S(\hat\rho_{\rm av})-S(\hat\rho)\ ,
\eeq
and so \eqr{eq: I_sym 1st version} can be written as
\beq
  I_{\rm sym}(\hat\rho)\le S({\cal G}[\hat\rho_{\rm av}])-S({\cal G}[\hat\rho])\ .
\eeq
Using \eqr{eq: max info in symmetry} then gives
\beq    \label{eq: I_sy bound}
  I_{\rm sym}(\hat\rho)\le \log D  -S({\cal G}[\hat\rho])\ .
\eeq
Comparing the right side with \eqr{eq: W_G} shows that the information
capacity $I_{\rm sym}(\hat\rho)$ is bounded by the entropic measure of the
symmetry $W_G(\hat\rho)$.

Moreover, the index $j$ associated with the state $\hat\rho_j$ that is sent
to Bob represents information about the symmetry of the state $\hat\rho$.
There is a total of $I_{\rm sym}(\hat\rho)$ bits of classical information
about the symmetry being sent to Bob and this is the maximum amount possible.
This implies that {\it Alice's knowledge of the original state $\hat\rho$
represents $I_{\rm sym}(\hat\rho)$ bits of classical information about the
symmetry.}

\section{Duality of symmetry and asymmetry }

\subsection{Duality relation}

Combining the two expressions (\ref{eq: I_as bound}) and (\ref{eq: I_sy
bound}) yields the duality relation,
\beq  \label{eq: duality sym asym}
   I_{\rm asym}(\hat\rho)+I_{\rm sym}(\hat\rho)\le \ln(D)-S(\hat\rho)\ ,
\eeq
which is the central result of this paper. It shows that the sum of the
information capacities of the symmetry and asymmetry is bounded by the
maximum classical information that can be carried by the system.

We could have anticipated this relation from the manner in which we designed
the encodings:  the encoding of $I_{\rm asym}(\hat\rho)$ bits of information
in the asymmetry of the state leaves the symmetry unchanged, and so a further
$I_{\rm sym}(\hat\rho)$ bits of information can subsequently be encoded in
the symmetry of the same system (and vice versa for the reverse order of
encodings). Indeed, the sets of operators, $\{\hat T_g$\} and $\{\hat U_j\}$,
used in the two encodings commute in the sense of \eqr{eq: G invariant op}.
The sum $I_{\rm asym}(\hat\rho)+I_{\rm sym}(\hat\rho)$, therefore, is {\it
necessarily} bounded by the maximum amount of classical information that can
be carried by the system.

But the importance of the inequality (\ref{eq: duality sym asym}) does not
lie in the actual value of the bound.  Rather, it is the fact (\ref{eq:
duality sym asym}) is a duality between information about the symmetry and
the asymmetry of the system. Knowing that the state of the system is
$\hat\rho$ gives us $I_{\rm asym}(\hat\rho)$ bits of information about its
asymmetry and $I_{\rm sym}(\hat\rho)$ bits of information about its symmetry.
Regardless of how the system is prepared, we can never have both $\log(D)$
bits of information about its symmetry and $\log(D)$ bits of information
about its asymmetry. Inequality (\ref{eq: duality sym asym}) therefore {\it
represents a duality between our knowledge of the asymmetry and the symmetry
of the system.}

In \S\ref{sec:symmetry} we expressed classical particle and wave character of
a quantum system in terms the asymmetry and symmetry, respectively, with
respect to a group. Inequality (\ref{eq: duality sym asym}) {\it represents a
duality between our knowledge of the particle and the wave character of the
system} in the same manner.

\subsection{Examples}

It is instructive to explore the duality relation (\ref{eq: duality sym
asym}) for states of particular interest. Consider a system which is capable
of completely breaking the symmetry of $G$ as discussed in
\S\ref{sec:symmetry}({\it\ref{sec:review sym and asym}}). Let the system be
prepared in one of the reference states defined in \eqr{eq: reference state
|g>}, say $\hat\rho=\ket{g}\bra{g}$ where $g\in G$. In this case the possible
density operators prepared by Alice are simply $\hat T_{g'} \hat\rho \hat
T_{g'}^\dagger=\ket{g'\circ g}\bra{g'\circ g}$ for $g'\in G$; these density
operators are mutually commuting and so the equality in Holevo's bound in
\eqr{eq: I_as bound} is satisfied (Ruskai 2002). Moreover $S(\hat\rho)=0$ and
the averaged state sent to Bob is proportional to the identity operator which
means $S({\cal G}[\hat\rho])=\log(D)$. The reference states therefore exhibit
the maximum information capacity possible, i.e.
\beq   \label{eq: I_as max}
   I_{\rm asym}(\ket{g}\bra{g})=\log(D)
\eeq
and so the bound in \eqr{eq: I_as bound} is achievable. As $S({\cal
G}[\hat\rho])=\log(D)$ we find from \eqr{eq: I_sy bound} that
\beq
   I_{\rm sym}(\ket{g}\bra{g})=0
\eeq
and so the symmetry has zero information capacity. Combining these results
shows that the equality in \eqr{eq: duality sym asym} is satisfied:
\beq
   I_{\rm asym}(\ket g\bra g)+I_{\rm sym}(\ket g\bra g)=\log(D)\ .
\eeq
Another way to state these results is that preparing the system in a
reference state gives maximum information about the asymmetry and minimum
information about the symmetry of the system.

There are no symmetric pure states for irreducible representations of
dimension greater than 1. In particular, a uniform linear superposition of
all the reference states of the form $\ket\phi\propto\sum_{g\in G} \ket g$
satisfies the symmetric condition ${\cal
G}\left[\ket\phi\bra\phi\right]=\ket\phi\bra\phi$, however, this state is
easily shown to be the trivial representation of dimension 1. For this state
we find
\beq
   I_{\rm sym}(\ket\phi\bra\phi)= \log(D)
\eeq
for an encoding in which Alice sends mutually commuting density operators
$\hat U_j\ket\phi\bra\phi\hat U_j^\dagger$ (Ruskai 2002).  As it is a
symmetric pure state, $S({\cal G}[\ket\phi\bra\phi])=0$ and so from \eqr{eq:
I_as bound}
\beq
   I_{\rm asym}(\ket\phi\bra\phi)= 0\ .
\eeq
Hence
\beq
   I_{\rm asym}(\ket\phi\bra\phi)+I_{\rm sym}(\ket\phi\bra\phi)=\log(D)
\eeq
and so the equality in \eqr{eq: duality sym asym} is satisfied. Preparing the
system in a pure symmetric state clearly gives maximum information about the
symmetry and no information about the asymmetry of the system.

\section{Composite systems \label{sec: composite systems}}

\subsection{Duality relation for composite system}

It is rather straightforward to extend the analysis above to a composite
system consisting of two identical systems, labelled $a$ and $b$. Let each
system be of the kind we have been considering, with a $D$-dimensional
Hilbert space $H$ that carries a representation of the group $G$.  The
Hilbert space $H\otimes H$ of the composite $ab$-system carries a
representation of $G$ which acts {\it globally} (or collectively) in the
sense that its unitary representation $\tau^{(ab)}=\{\hat T_g^{(ab)}: g\in
G\}$, where
\beq  \label{eq: T_g collection}
   \hat T_g^{(ab)}\equiv\hat T_g\otimes \hat T_g\ ,
\eeq
applies the same group action to each system (Fulton \& Harris 1991).  Here
$\hat A\otimes\hat B$ represents a tensor product of the operators $\hat A$
and $\hat B$ which act on the $a$ and $b$ systems, respectively.

It is straightforward to show that the analysis of the previous section with
respect to the group representation $\tau^{(ab)}$ on $H\otimes H$ gives the
bound on the information capacity $I^{(ab)}_{\rm asym}$ of the composite
system for an encoding in terms of the operators $\hat T_g^{(ab)}$ as
\beq  \label{eq: I_as collection}
   I_{\rm asym}^{(ab)}(\hat \rho)\le S({\cal G}^{(ab)}[\hat \rho])-S(\hat\rho)
\eeq
for the state $\hat\rho$ of the $ab$-system with
\beq  \label{eq: twirl collection}
   {\cal G}^{(ab)}[\hat\rho]\equiv\frac{1}{|G|}\sum_{g\in G} \hat T_g^{(ab)} \hat\rho \hat
   T_g^{(ab)}{}^\dagger\ .
\eeq
Similarly, the bound on the information capacity of an encoding in terms of
the symmetry of $\hat\rho$ is
\beq  \label{eq: I_sy collection}
   I^{(ab)}_{\rm sym}(\hat \rho)\le \log(D^2)-S({\cal G}^{(ab)}[\hat \rho])\ .
\eeq
Hence the duality relation for the $ab$-system is
\beq   \label{eq: duality collection}
   I^{(ab)}_{\rm asym}(\hat \rho)+I^{(ab)}_{\rm sym}(\hat \rho)\le 2\log(D)-S(\hat \rho) \ .
\eeq
The $ab$-system has the capacity to carry twice the classical information as
each single system and this is reflected in the larger bound here compared to
\eqr{eq: duality sym asym}.

\subsection{Examples}

It is interesting to compare the bounds on the symmetry and asymmetry for a
composite of two systems each of which is capable of completely breaking the
symmetry of a finite group $G$. The product reference state $\ket
g\otimes\ket g$ gives the maximum information capacity of the asymmetry of
the $ab$-system of
\beq
\label{eq: I_as max collection}
    I_{\rm asym}^{(ab)}(\ket g\otimes\ket g)=\log(D)\ .
\eeq
Here, and in the following, we write the state as a simple Dirac ket rather
than a density operator to make the notation easier. It might have been
expected that the asymmetry would be twice this amount because the sum of the
asymmetry for each system in \eqr{eq: I_as max} is $2\log(D)$. However,
$I_{\rm asym}^{(ab)}$ is the global asymmetry of the $ab$-system for which
$G$ acts identically on the individual systems according to \eqr{eq: T_g
collection}. The information capacity of the symmetry is easily found to be
\beq
    I_{\rm sym}^{(ab)}(\ket{g}\otimes\ket{g})=\log(D)
\eeq
and so the product reference state is as asymmetric as it is symmetric.
Conversely, the state $\ket{\varphi}=\sum_{g\in
G}\ket{g}\otimes\ket{g}/\sqrt{n_G}$ has the minimum asymmetry and maximum
symmetry:
\beq
    I_{\rm asym}^{(ab)}(\ket{\varphi})=0, \quad
    I_{\rm sym}^{(ab)}(\ket{\varphi})=2\log(D)\ .
\eeq

Although the state $\ket{\varphi}$ has no (global) asymmetry, nevertheless it
does possess {\it local} asymmetry in the sense that it is transformed by the
action of $\hat T_g\otimes\hat T_{h}$ for particular choices of $g, h\in G$
to a different state. Indeed the transformed state
\beq  \label{eq: local asym}
   \hat T_g\otimes\hat T_{h}\ket{\varphi}=\hat T_e\otimes\hat
   T_{r}\ket{\varphi}
\eeq
is orthogonal to $\ket{\varphi}$ unless $r=e$, where $r=h \circ g^{-1}$ and
$e$ is the identity element. Thus while the composite system as a whole is
symmetric, it has asymmetric parts internally. Moreover, the transformed
state in \eqr{eq: local asym} is also globally symmetric.  Evidently the
asymmetry of one component system is compensated by the asymmetry of the
other. The state $\ket{\varphi}$ is maximally entangled and its symmetry is
due to it belonging to a 1 dimensional representation of $G$ on $H\otimes H$.
This situation is related to the superdense coding scheme of Bennett \&
Wiesner (1992) which we examine in the next section.

These results are easily extended to the general case of $n$ systems as
follows:
\beq
    I^{(ab\cdots)}_{\rm asym}(\ket{g}^{\otimes n})=\log(D)\ ,\quad
    I^{(ab\cdots)}_{\rm sym}(\ket{g}^{\otimes n})=(n-1)\log(D)
\eeq
whereas
\beq
    I^{(ab\cdots)}_{\rm asym}(\frac{1}{\sqrt{n_G}}\sum_{g}\ket{g}^{\otimes n})=0\ ,\quad
    I^{(ab\cdots)}_{\rm sym}(\frac{1}{\sqrt{n_G}}\sum_{g}\ket{g}^{\otimes n})=n\log(D)\ .
\eeq
Clearly more systems implies greater symmetry and thus more wavelike
character. These results stem directly from the manner in which the group
acts globally, according to \eqr{eq: T_g collection}.  For the cases
considered here, that is for finite groups and systems that can completely
break the symmetry according to \eqr{eq: reference state |g>}, the maximum of
$I^{(ab\cdots)}_{\rm asym}$ remains fixed at $\log(D)=\log(n_G)$ as the
number $n$ of systems grows.  The total amount of classical information that
can be carried by the composite system is $n\log(D)$, and so the maximum
amount of information that can be carried by the symmetry also scales
linearly with $n$.

\section{Applications}

\subsection{Two-path interferometer}

\begin{figure}  
\begin{center}
\includegraphics[width=8cm,height=5.1cm]{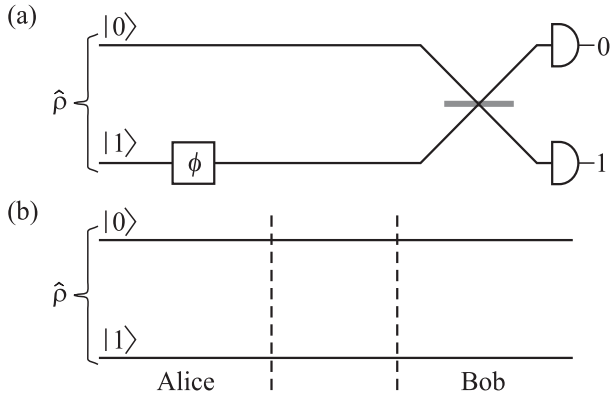}
\end{center}
\caption{Quantum circuit diagrams of (a) a two path interferometer and (b)
its representation as a communication channel. A wavicle enters from the left
in a given state $\hat\rho$. In (a) interference with respect to the phase
shift $\phi$ is observed in the outputs of the detectors on the right. In (b)
Alice can encode information in the symmetry of asymmetry of $\hat\rho$. The
maximum amount of information that can be transmitted to Bob in this way is
the information Alice has about the symmetry or asymmetry from her knowledge
of $\hat\rho$. \label{fig2: interferometer}}
\end{figure}    

As an application of the single system case let's examine a two-path
interferometer. A quantum system, or ``wavicle'' if you will (Eddington
1928), enters the left side of the interferometer illustrated in figure
\ref{fig2: interferometer}(a) in the state $\hat\rho$, passes through the
beam splitter and is detected by two detectors on the right. Let the states
representing the wavicle occupying the upper and lower paths of the
interferometer be $\ket 0$ and $\ket 1$, respectively. A phase shift $\phi$
is applied to the lower path as shown. The action of the phase shifter and
beam splitter on the state of the wavicle is described by $\hat\rho \mapsto
\hat U(\phi)\hat\rho \hat U^\dagger(\phi)$ where  $\hat
U(\phi)=\frac{1}{\sqrt{2}}(\hat\id+\ri\hat\sigma_y)\exp(\ri \phi \ket 1\bra
1)$ and $\hat \sigma_y=\ri(\ket{1}\bra{0}-\ket{0}\bra{1})$. The probability
that the upper detector detects the wavicle depends on $\phi$ and represents
an interference pattern whose visibility is given by
$V=2|\ip{1|\hat\rho|0}|$.

The symmetry-asymmetry duality arises when we treat the wavicle--two-path
system as a communication channel between two parties, Alice and Bob, as
illustrated in figure \ref{fig2: interferometer}(b). The interchange of the
two paths of the communication channel is the two-state equivalent of the
continuous spatial translation discussed in
\S\ref{sec:symmetry}({\it\ref{sec:part and wave props}}). Just as we defined
the symmetry of a classical wave in \S\ref{sec:symmetry}({\it\ref{sec:part
and wave props}}) in terms of the invariance to spatial translations, here
symmetry is defined in terms of the invariance to path interchange. The
symmetry group associated with this problem is therefore given by $G=\{e,x\}$
where $e$ is the identity element and $x$ represents the interchange of the
paths.

A suitable unitary representation of $G$ is given by $\tau =\{\hat\id,\hat
\sigma_x\}$ where $\hat\id$ is the identity operator and $\hat \sigma_x=\ket
1\bra 0+\ket 0\bra 1$.  Alice can encode information in the symmetry or
asymmetry of $\hat\rho$ with respect to $G$ and transmit this information to
Bob via the wavicle.  In particular, the information capacity of the
asymmetry, $I_{\rm asym}(\hat\rho)$, represents classical information about
the asymmetry that Alice has from her knowledge that the state is $\hat\rho$.
This information represents what she knows about the behaviour of the wavicle
under the action of the operator $\hat \sigma_x$ which interchanges the
paths.  In other words, it represents what she knows about the
distinguishability of the paths from her knowledge of $\hat\rho$ and so
$I_{\rm asym}(\hat\rho)$ plays essentially the same role as Englert's
which-way information (Englert 1996).

Consider first the states $\ket{n}$ where $n=0$ or $1$. These represent the
wavicle occupying one of the paths and so they are particle-like states.
From equations (\ref{eq: I_as bound}), (\ref{eq: I_sy bound}) and (\ref{eq:
duality sym asym}) we find
\beq
   I_{\rm asym}(\ket n)=1\ ,\quad I_{\rm sym}(\ket n)=0\ ,
   \quad I_{\rm asym}(\ket n)+I_{\rm sym}(\ket n)=\log(D)
\eeq
which shows that the state is maximally asymmetric or particle like. The same
states would give a visibility $V$ of zero in the interferometer in figure
\ref{fig2: interferometer}(a).

Next consider the states representing the wavicle occupying an equal
superpositions of the two paths, $\ket{\pm}=(\ket{0}\pm\ket{1})/\sqrt{2}$. In
this case we find
\beq
   I_{\rm asym}(\ket\pm)=0\ ,\quad I_{\rm sym}(\ket\pm)=1\ ,
   \quad I_{\rm asym}(\ket\pm)+I_{\rm sym}(\ket\pm)=\log(D)
\eeq
which shows that the state is maximally symmetric or wave like. The same
states would give a visibility $V$ of unity in the interferometer.  Evidently
the symmetry of the system with respect to the group $G$ encompasses the
classical wave-like properties that are probed by the interferometer.

We argued in \S\ref{sec:symmetry}({\it\ref{sec:sym and interference}}) that
interferometry can be viewed as a particular communication protocol that
encodes information in the symmetry of a system. Let's now examine this idea
in the context of the two-path interferometer in figure \ref{fig2:
interferometer}(a). Imagine that Alice uses the identity and the phase shift
operators to encode information using the initial state $\ket{+}$ and Bob
uses the beam splitter and detector arrangement for decoding.  We saw in
\S\ref{sec: info capacity}({\it \ref{subsec: Info capacity - sym}}) that for
the encoding to use only the symmetry of the system it must not change the
asymmetry.  This is clearly the case as the initial state $\ket{+}$ and the
possible encoded states $\ket{\pm}$ all have zero asymmetry as does the
average encoded state.  But we should also check that the encoding operators
are $G$-invariant in the sense of \eqr{eq: G invariant op}.  For this, first
note that the Pauli operators satisfy
\beq  \label{eq: sigma_x sigma_y = ie sigma_z}
  \hat\sigma_i\hat\sigma_j=\ri e_{ijk}\hat\sigma_k\quad \mbox{(for $i$, $j$ and $k$ all
  different)}
\eeq
where $e_{ijk}$ is the permutation symbol and $\hat\sigma_1=\hat\sigma_x$,
$\hat\sigma_2=\hat\sigma_y$ and $\hat\sigma_3=\hat\sigma_z$.  It immediately
follows that
\beq
   \hat\sigma_x(\hat\sigma_z)^n\hat\rho(\hat\sigma_z)^n\hat\sigma_x
   =(\hat\sigma_z)^n\hat\sigma_x\hat\rho\hat\sigma_x(\hat\sigma_z)^n
\eeq
and so
\beq
   \hat\sigma_x f(\hat\sigma_z)\hat\rho f(\hat\sigma_z)\hat\sigma_x
   = f(\hat\sigma_z)\hat\sigma_x\hat\rho \hat\sigma_x f(\hat\sigma_z)
\eeq
for any function $f(\cdot)$ with a power series expansion. This means that
the phase shift operator $\exp(\ri \phi \ket 1\bra 1)=\exp[\ri
\phi(\hat\id-\hat\sigma_z)]$ is indeed $G$-invariant.  As the identity
operator is trivially $G$-invariant, we conclude that the two-path
interferometer can act as a communication channel that encodes information in
the symmetry only, as anticipated.

\subsection{Superdense coding}

\begin{figure}  
\begin{center}
\includegraphics[width=8cm,height=5.0cm]{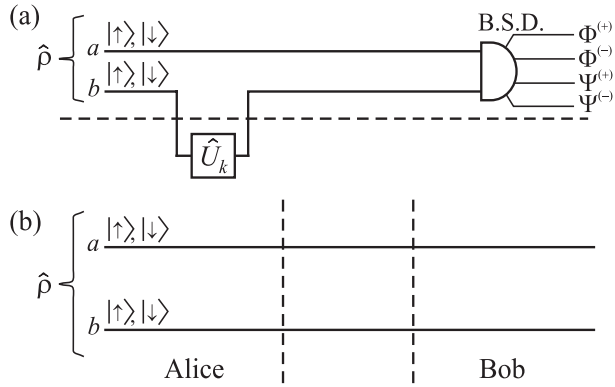}
\end{center}
\caption{Quantum circuit diagrams of (a) the superdense coding scheme and (b)
its representation as a symmetric and asymmetric encoding communication
channel. A pair of spin-$\frac{1}{2}$ particles labelled $a$ and $b$ enters
from the left in a given state $\hat\rho$. In (a) a local operator $\hat
U_k\in\{\hat\id,\hat\sigma_x, \hat \sigma_y,\hat\sigma_z\}$ is applied to the
lower spin and then both spins pass to the Bell state discriminator (B.S.D.).
In (b) Alice can encode information using the symmetry or asymmetry of
$\hat\rho$ and send it to Bob. \label{fig3: superdense}}
\end{figure}    

The superdense coding scheme of Bennett \&  Wiesner (1992) allows 2 bits of
classical information to be communicated using a single spin-$\frac{1}{2}$
particle and pre-existing entanglement. Figure \ref{fig3: superdense}(a)
illustrates the scheme.  A pair of spin-$\frac{1}{2}$ particles labelled $a$
and $b$ are prepared in the Bell state $\ket{\Psi^{(-)}}$. Here the four Bell
states are given by
$\ket{\Phi^{(\pm)}}=(\ket{\up}\ket{\up}\pm\ket{\dn}\ket{\dn})/\sqrt{2}$ and
$\ket{\Psi^{(\pm)}}=(\ket{\up}\ket{\dn}\pm\ket{\dn}\ket{\up})/\sqrt{2}$ where
$\ket{\up}$ and $\ket{\dn}$ represent eigenstates of the $z$ component of
spin with eigenvalues $\hbar/2$ and $-\hbar/2$, respectively (Braunstein {\it
et al.} 1992). A local operation $\hat U_k\in\{\hat\id,\hat\sigma_x, \hat
\sigma_y,\hat\sigma_z\}$, where $\hat\sigma_x$, $\hat\sigma_y$ and
$\hat\sigma_z$ are Pauli spin operators, applied to the lower spin transforms
the state of the pair into one of the Bell states.  The value of $k$ can be
determined by a detector which discriminates between these four states.

Although we have used the Bell state $\ket{\Psi^{(-)}}$ in this example, the
same coding and detection scheme in figure \ref{fig3: superdense}(a) works
with the system initially prepared in any of the four Bell states. Superdense
coding can also work with any maximally entangled pure state, but in that
case the coding and detection methods in figure \ref{fig3: superdense}(a)
need to be modified accordingly. For brevity, we shall only treat the
encoding and detection scheme depicted explicitly in figure \ref{fig3:
superdense}(a).

We now reexamine the scheme using the information capacities of the symmetry
and asymmetry for a composite system. We first need to identify the symmetry
group associated with the problem.  The symmetry operations will be analogous
to the spatial translations mentioned in
\S\ref{sec:symmetry}({\it\ref{sec:part and wave props}}) and the path
exchange for the previous example. Specifically, we need to look for those
operations that leave the initial prepared state unchanged in the sense of
\eqr{eq: G rho=rho}, which for the composite system here is
\beq  \label{eq: G[rho]=rho collection}
    {\cal G}^{(ab)}[\hat\rho]= \hat\rho\ ,
\eeq
where ${\cal G}^{(ab)}[\cdot]$ is given by equations (\ref{eq: T_g
collection}) and (\ref{eq: twirl collection}). One might be tempted to say
that the operations we want are the group of global rotations $\{\hat
T_g\otimes\hat T_g:\hat T_g\in \mbox{SU(2)}\}$ because these operations leave
the singlet state $\ket{\Psi^{(-)}}$ invariant. However, any of the four Bell
states can be used in the same experimental arrangement in figure \ref{fig3:
superdense}(a) whereas only the singlet state is invariant under this group.
The symmetry described by this group is, therefore, not sufficiently general.
Instead, it is straightforward to show that a subgroup comprising the global
Pauli operators, i.e.
\beq  \label{eq: global Pauli ops}
   G=\{\hat\id\otimes\hat\id,\hat\sigma_x\otimes\hat\sigma_x,
      \hat\sigma_y\otimes\hat\sigma_y,\hat\sigma_z\otimes\hat\sigma_z\}\ ,
\eeq
leaves all four Bell states invariant in the sense of \eqr{eq: G[rho]=rho
collection}.  We therefore take this group to define the symmetry of the
problem.

The maximum value of asymmetry for pure states occurs when ${\cal
G}^{(ab)}[\hat\rho]$ is proportional to the identity operator. An example of
a state with this property is
$\ket{\phi}=(\ket{\up}\otimes\ket{\up}+\ket{\up}\otimes\ket{\dn})/\sqrt{2}$
$=(\ket{\Phi^{(+)}}+\ket{\Phi^{(-)}}+\ket{\Psi^{(+)}}+\ket{\Psi^{(-)}})/2$
for which we find, using equations (\ref{eq: I_as
 collection}), (\ref{eq: I_sy collection}) and (\ref{eq: duality
 collection}), that
\beq
   I^{(ab)}_{\rm asym}(\ket{\phi})=2\ , \quad I^{(ab)}_{\rm sym}(\ket{\phi})=0\ ,
   \quad I^{(ab)}_{\rm asym}(\ket{\phi})+I^{(ab)}_{\rm sym}(\ket{\phi})=2\log(D)
\eeq
where $D=2$ is the dimension of the Hilbert space of each spin.  The bound of
the duality \eqr{eq: duality collection} is reached for zero information
about the symmetry, and so this state is maximally asymmetric.

In contrast, all Bell states
$\ket{B}\in\{\ket{\Phi^{(+)}},\ket{\Phi^{(-)}},\ket{\Psi^{(+)}},\ket{\Psi^{(-)}}\}$
are found to satisfy
\beq
   I^{(ab)}_{\rm asym}(\ket{B})=0\ , \quad I^{(ab)}_{\rm sym}(\ket{B})=2\ ,
   \quad I^{(ab)}_{\rm asym}(\ket{B})+I^{(ab)}_{\rm sym}(\ket{B})=2\log(D)
\eeq
and so they have maximal symmetry.  In particular, 2 bits of classical
information can encoded in their symmetry.

It remains for us to show that the superdense coding scheme is described by
the formalism of \S\ref{sec: composite systems}.  The operators used in the
superdense coding scheme are are given by
\beq
   U=\{\hat\id\otimes\hat\id,\hat\id\otimes\hat\sigma_x,\hat\id\otimes\hat\sigma_y,
     \hat\id\otimes\hat\sigma_z\}
\eeq
and it is straightforward to show using \eqr{eq: sigma_x sigma_y = ie
sigma_z} that they are $G$-invariant with respect to $G$ defined in \eqr{eq:
global Pauli ops}. The initial state of the scheme can be any of the bell
states $\ket{B}$ and these have no asymmetry and neither do any of the
possible encoded states.  The superdense coding scheme therefore encodes the
information in the symmetry of the spin particles in accord with \S\ref{sec:
composite systems}.

Comparing this with the previous example shows that one can view superdense
coding as being essentially an {\it interference phenomena} involving
wave-like properties with respect to the group $G$ in \eqr{eq: global Pauli
ops}.  In this view the different outputs at the Bell-state discriminator
correspond to interference fringes.

Although all Bell states $\ket{B}$ are globally symmetric with respect to $G$
they are asymmetric in the local sense. In fact the operators $\hat U_k\in U$
used in the superdense coding scheme can be written as $\hat T_e\otimes\hat
T_g$ with $e$ being the identity element and $\hat T_g
\in\{\hat\id,\hat\sigma_x, \hat \sigma_y,\hat\sigma_z\}$. This encoding is in
the form of \eqr{eq: local asym}.  Thus the 2 bits of classical information
is encoded in the {\it local} asymmetry of $\ket{B}$.

\section{Discussion}

We began with the observation that a classical point particle breaks
translational symmetry whereas a classical wave of uniform amplitude does
not. A quantum system can be in a superposition of many positions which gives
the system varying degrees of translational symmetry. The superposition state
also allows the system to produce interference patterns in an interferometer
and thus to exhibit wave-like character.  The more translational symmetry the
system has, the more wave like it is. Conversely, the less translational
symmetry the system has, the more particle like it is. The broad aim of this
paper is to elevate these associations to formal definitions of particle and
wave nature for arbitrary quantum systems and arbitrary symmetry groups. To
do this we quantified the symmetry and asymmetry of a system with respect to
an arbitrary group $G$ in terms of the amount of information each can carry.
This led to the particle-wave duality relations for a single system (\ref{eq:
duality sym asym}) and for a composite system (\ref{eq: duality collection}).

We then used the duality relations to recast the well-known concepts of
two-path interferometry and superdense coding in new light. In both cases the
first task in applying the duality relations was to find the symmetry group
associated with the initial states of the two schemes. This entailed finding
operations which were analogous to the spatial translations that leave a
classical wave invariant.  The operations found were path exchange for the
two-path interferometer and global Pauli operators for the superdense coding
scheme. We showed how the two-path interferometer can be viewed as a
communication channel which encodes information in the symmetry of the
initial state.  Conversely, we showed how the superdense coding scheme can be
viewed as an interferometer whose operation is to use the wave-like symmetry
of its initial state to produce outputs that correspond to interference
fringes.

The outcomes of this study are fourfold. First, it shows how a communication
channel based on symmetry can be viewed as an interferometer, and vice versa.
As an interferometer demonstrates the degree to which a system has wave-like
properties, it shows how arbitrary symmetries can be interpreted in terms of
wave-like and particle-like character. Second, it also establishes the
duality relation that governs the amount of information that is available
about each character from knowledge of the state $\hat\rho$ of the system.
Third, in doing so it replaces pairs of conjugate observables, such as
position and momentum, with {\it sets of operators} associated with the
classical notions of particle and wave. The set of operators belonging to the
group $G$ manipulate the particle-like asymmetry, and the set of operators
that are $G$-invariant manipulate the wave-like symmetry of the system.
Fourth, this approach avoids the highly mathematical nature of the mutually
unbiased bases method by offering a way of studying particle-way duality that
is {\it general and yet retains a consistent physical interpretation}. These
outcomes answer the questions posed in the Introduction.

Finally, we end with a connection between the duality studied here and
complementarity. Although we have not made explicit reference to it, there
must exist some prior measurement on the system in order to be able to know
anything about its state. As pointed out by Bohr, measurement schemes that
reveal unambiguous information about complementary observables are mutually
exclusive (1935). Our knowledge of $\hat\rho$ is therefore subject to the
same limitations of complementarity. These limitations are implicit in the
duality relations as limits on what can be known about the system in terms of
its symmetry and asymmetry.

\ack{I would like to thank Prof. Howard Wiseman for encouraging discussions.
This research was conducted by the Australian Research Council Centre of
Excellence for Quantum Computation and Communication Technology (Project
number CE110001027).
}

\label{lastpage}

\end{document}